\documentclass[prd,twocolumn,groupedaddress,showpacs,nofootinbib]{revtex4}
\usepackage{graphicx}
\usepackage{dcolumn}
\usepackage{amssymb}
\usepackage{mathrsfs}
\usepackage{amsmath}
\usepackage{epsfig}
\usepackage[dvips]{color}
\usepackage{hhline}

\begin{document}

\title{Double type-II Dirac seesaw accompanied by Dirac fermionic dark matter}

\author{Pei-Hong Gu}
\email{peihong.gu@sjtu.edu.cn}

\affiliation{School of Physics and Astronomy, Shanghai Jiao Tong University, 800 Dongchuan Road, Shanghai 200240, China}

\begin{abstract}

A TeV-scale Higgs doublet with a small mixing to the standard model Higgs doublet can have the sizable Yukawa couplings to several right-handed neutrinos and the standard model lepton doublets. This provides a testable Dirac neutrino mass generation. We further consider a seesaw mechanism involving a $U(1)_{B-L}^{}$ gauge symmetry, which predicts the existence of two right-handed neutrinos and a stable Dirac fermionic dark matter, to simultaneously explain the small mixing between the two Higgs doublets and the generation of the cosmic baryon asymmetry.

\end{abstract}


\maketitle

\section{Introduction}

The concept of three massive and mixing neutrinos has been established by the atmospheric and solar neutrino oscillations and then has been confirmed by the accelerator and reactor neutrino oscillations \cite{pdg2018}. Meanwhile, the cosmological observations have constrained the neutrinos to be extremely light \cite{pdg2018}. Currently, various seesaw \cite{minkowski1977} mechanisms at tree or loop level \cite{minkowski1977,mw1980,flhj1989,ma2006} are considered the best explanation for the tiny but nonzero neutrino masses. In this popular seesaw scenario, the neutrino masses are induced by certain lepton-number-violating interactions so that the neutrinos should have a Majorana nature. However, the lepton number violation and then the Majorana neutrinos are just a theoretical assumption and have not been seen in any experiments. So, we should also consider the possibility of Dirac neutrinos \cite{rw1983,dlrw1999,mp2002,tt2006,gh2006,gu2016,gs2007,gu2012,gu2017,gu2019-1,es2015,bd2016,aas2017,wh2017,yd2017,csv2018,cryz2019,bdhps2019,saad2019,ekss2019,jks2019,ma2019}. Actually, the so-called type-I \cite{rw1983}, type-II \cite{gh2006} and type-III \cite{gu2016} Dirac seesaw as well as the radiative Dirac seesaw \cite{gs2007} for the Dirac neutrinos have been constructed in analogy to the conventional seesaw mechanisms for the Majorana neutrinos. In the Majorana or Dirac seesaw models, the cosmic baryon asymmetry, which is another big challenge to the standard model (SM), can be understood in a natural way \cite{dlrw1999,mp2002,tt2006,gh2006,gu2016,gs2007,gu2012,gu2017,gu2019-1,fy1986,lpy1986,fps1995,ms1998,bcst1999,hambye2001,di2002,gnrrs2003,hs2004,bbp2005,ma2006}. This is the famous leptogenesis mechanism \cite{fy1986}.

In a recent work \cite{gu2019-1}, we have proposed a double Dirac seesaw model where a TeV-scale Higgs doublet with a small mixing to the SM Higgs doublet can have the sizable Yukawa couplings to several right-handed neutrinos and the SM lepton doublets. This Dirac neutrino mass generation could be tested experimentally \cite{wwy2006,dl2010}. We showed the small mixing between the two Higgs doublets could come from certain interactions for generating the cosmic baryon asymmetry.

In this paper we shall develop our idea of the double Dirac seesaw to naturally include a dark matter particle, which is absent from the SM. Specifically we shall introduce a $U(1)_{B-L}^{}$ gauge symmetry with four right-handed neutrinos. Through their Yukawa couplings to a Higgs singlet for spontaneously breaking the $U(1)_{B-L}^{}$ symmetry, two right-handed neutrinos can form a Dirac fermion to be a stable dark matter particle. As for the other two right-handed neutrinos, they can couple to the SM lepton doublets with a second Higgs doublet. After the $U(1)_{B-L}^{}$ symmetry breaking, two or more heavy Higgs singlets can acquire their suppressed vacuum expectation values (VEVs) to result in a small mixing between the second Higgs doublet and the SM Higgs doublet. Three left-handed neutrinos and two right-handed neutrinos thus can have a mass matrix with two nonzero eigenvalues. The heavy Higgs singlet decays can generate an asymmetry stored in the second Higgs doublet. This asymmetry can lead to a baryon asymmetry in association with the sphaleron processes \cite{krs1985}.

\section{Fermions and scalars}

Under the $SU(3)_c^{} \times SU(2)^{}_{L}\times U(1)_Y^{}\times U(1)_{B-L}^{}$ gauge symmetries, we need extend the SM fermions,
\begin{eqnarray}
\label{sm}
&&\begin{array}{l}q^{}_{L}(3,2,+\frac{1}{6})(+\frac{1}{3})\,,\end{array} ~~\begin{array}{l}d^{}_{R}(3,1,-\frac{1}{3})(+\frac{1}{3})\,,\end{array} \nonumber\\
[1mm]
 &&\begin{array}{l}u^{}_{R}(3,1,+\frac{2}{3})(+\frac{1}{3})\,,\end{array} ~~\begin{array}{l}l^{}_{L}(1,2,-\frac{1}{2})(-1)\,,\end{array} \nonumber\\
[1mm]
&&\begin{array}{l}e^{}_{R}(1,1,-1)(-1)\,,\end{array} 
\end{eqnarray} 
by some right-handed neutrinos to cancel the $U(1)_{B-L}^{}$ gauge anomalies \cite{mp2007,pry2016,gu2019-2}. Here and thereafter the first and second brackets following the fields respectively describe the transformations under the $SU(3)_c^{} \times SU(2)^{}_{L}\times U(1)_Y^{}$ gauge groups and the $U(1)_{B-L}^{}$ gauge group. In the present work, we consider the following four right-handed neutrinos,
\begin{eqnarray}
&&\begin{array}{l}\nu^{}_{R1,2}(1,1,0)(-\frac{8}{5})\,,\end{array} \begin{array}{l}\nu^{}_{R3}(1,1,0)(\frac{1+\sqrt{865}}{10})\,,\end{array} \nonumber\\
&& \begin{array}{l}\nu^{}_{R4}(1,1,0)(\frac{1-\sqrt{865}}{10})\,.\end{array}
\end{eqnarray}

While the SM Higgs doublet
\begin{eqnarray}
\begin{array}{l}\phi(1,2,-\frac{1}{2})(0)\,,\end{array} 
\end{eqnarray} 
is responsible for the spontaneous electroweak symmetry breaking, we introduce a Higgs singlet, 
\begin{eqnarray}
\begin{array}{l}
\xi(1,1,0)(+\frac{1}{5}),\end{array}
\end{eqnarray}
to drive the spontaneous $U(1)_{B-L}^{}$ symmetry breaking.

It is easy to see the Higgs singlet $\xi$ can have a Yukawa interaction with the third and forth right-handed neutrinos $\nu_{R3,4}^{}$, i.e.
\begin{eqnarray}
\label{dm}
\mathcal{L}\supset - y_{34}^{} \left(\xi \bar{\nu}_{R3}^{} \nu_{R4}^{c} +\textrm{H.c.}\right)\,.
\end{eqnarray}
Furthermore, in association with the Higgs singlet $\xi$, we can construct the following dimension-5 operators involving the SM lepton and Higgs doublets as well as the first and second right-handed neutrinos $\nu_{R1,2}^{}$, i.e.
\begin{eqnarray}
\label{numass}
\mathcal{L}\supset - \sum_{i=1,2 \atop \alpha=e,\mu,\tau}\frac{c_{ \alpha i}^{}}{\Lambda^3_{}} \bar{l}_{L\alpha}^{} \phi \nu_{Ri}^{}\xi^3_{} +\textrm{H.c.}\,.
\end{eqnarray}
The above effective operators can originate from two renormalizable models as below, 
\begin{eqnarray}
\label{laga}
\mathcal{L}_{A}^{}\!\!&\supset&\!\! - \left(m_\eta^2 +\lambda_{\eta\xi}^{}\xi^\dagger_{}\xi+  \lambda_{\eta\phi}^{}\phi^\dagger_{}\phi\right)\eta^\dagger_{}\eta - \lambda'^{}_{\eta\phi}\eta^\dagger_{}\phi \phi^\dagger_{}\eta\nonumber\\
\!\!&&\!\!-\sum_{a=1}^{n\geq 2}\left(M_{\sigma_a}^2 \sigma^\dagger_{a}\sigma_{a}^{} + \rho_{\sigma_a}^{} \sigma_a^{} \eta^\dagger_{}\phi + \kappa_{\sigma_a}^{}\sigma_a^{}\xi^\dagger_{}\xi^\dagger_{}\xi^\dagger_{}\right)\nonumber\\
\!\!&&\!\!  - \sum_{ i=1,2 \atop \alpha=e,\mu,\tau }^{} f_{\alpha i}^{} \bar{l}_{L\alpha}^{} \eta \nu_{Ri}^{} + \textrm{H.c.}\,,
\end{eqnarray}
\begin{eqnarray}
\label{lagb}
\mathcal{L}_{B}^{}\!\!&\supset&\!\!  - \left(m_\eta^2 +\lambda_{\eta\xi}^{}\xi^\dagger_{}\xi+  \lambda_{\eta\phi}^{}\phi^\dagger_{}\phi\right)\eta^\dagger_{}\eta - \lambda'^{}_{\eta\phi}\eta^\dagger_{}\phi \phi^\dagger_{}\eta\nonumber\\
\!\!&&\!\!-\sum_{b=1}^{n\geq 2}\left(M_{\omega_b}^2 \omega^\dagger_{b}\omega_{b}^{} + \kappa_{\omega_b}^{} \omega_b^{} \xi \eta^\dagger_{}\phi + \rho_{\omega_b}^{}\omega_b^{}\xi^\dagger_{}\xi^\dagger_{} \right)\nonumber\\
\!\!&&\!\! - \sum_{ i=1,2 \atop \alpha=e,\mu,\tau }^{} f_{\alpha i}^{} \bar{l}_{L\alpha}^{} \eta \nu_{Ri}^{} + \textrm{H.c.}\,,
\end{eqnarray}
where $\eta$ is a second Higgs doublet,
\begin{eqnarray}
\begin{array}{l}
\eta(1,2,-\frac{1}{2})(+\frac{3}{5})\,,\end{array}
\end{eqnarray}
while $\sigma_a^{}$ and $\omega_b^{}$ are two types of heavy Higgs singlets,  
\begin{eqnarray}
\begin{array}{l}\sigma_a^{}(1,2,-\frac{1}{2})(+\frac{3}{5})\,,\end{array}\begin{array}{l}\omega_b^{}(1,2,-\frac{1}{2})(+\frac{2}{5})\,. \end{array}
\end{eqnarray}
Note the model (\ref{laga}) or (\ref{lagb}) can always contains a relative phase among the four parameters $(\kappa_{\sigma_{a}}^{},\kappa_{\sigma_{c\neq a}}^{},\rho_{\sigma_{a}}^{},\rho_{\sigma_{c\neq a}}^{})$ or $(\kappa_{\omega_{b}}^{},\kappa_{\omega_{d\neq b}}^{},\rho_{\omega_{b}}^{},\rho_{\omega_{d\neq b}}^{})$.

\section{Dirac neutrino mass}

\begin{figure*}
\vspace{7.5cm} \epsfig{file=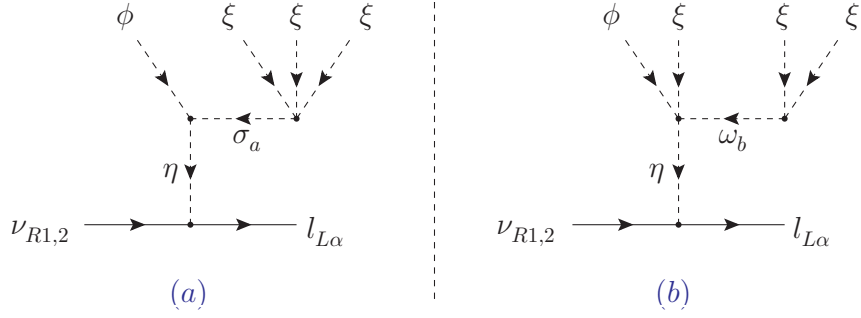, bbllx=5cm, bblly=6.0cm,
bburx=15cm, bbury=16cm, width=8cm, height=8cm, angle=0,
clip=0} \vspace{-10.5cm} \caption{\label{numass} The Dirac neutrino mass generation.}
\end{figure*}

As shown in Fig. \ref{numass}, we can obtain the effective operators (\ref{laga}) by integrating out the heavy Higgs singlets $\sigma_a^{}/\omega_b^{}$ and the second Higgs doublet $\eta$, i.e.
\begin{eqnarray}
\frac{c_{\alpha i}^{}}{\Lambda^3_{}} = -\sum_{a/b}^{}\frac{\rho_{\sigma_a/\omega_b}^{}\kappa_{\sigma_a/\omega_b}^{} }{M_{\sigma_a/\omega_b}^2 m_\eta^2} f_{\alpha i}^{} \,.
\end{eqnarray}
The three left-handed neutrinos $\nu_{L\alpha}^{}$ and the two right-handed neutrinos $\nu_{R1,2}^{}$ thus can acquire a tiny Dirac mass term after the $SU(2)_L^{}$ and $U(1)_{B-L}^{}$ symmetry breaking,
\begin{eqnarray}
\mathcal{L}&\supset& -\sum_{i=1,2 \atop \alpha=e,\mu,\tau}^{}\left(m_\nu^{}\right)_{\alpha i}^{} \bar{\nu}_{L\alpha}^{} \nu_{Ri}^{} +\textrm{H.c.}~~\textrm{with}\nonumber\\
&&\left(m_\nu^{}\right)_{\alpha i}^{} = -\sum_{a/b}^{}\frac{\rho_{\sigma_a/\omega_b}^{}\kappa_{\sigma_a/\omega_b}^{} \langle\xi\rangle^3_{} }{M_{\sigma_a/\omega_b}^2 m_\eta^2} f_{\alpha i}^{} \langle\phi\rangle\,.
\end{eqnarray} 
This Dirac neutrino mass generation actually is a two-step seesaw mechanism. We demonstrate the details in the following.

When the Higgs singlet develops its VEV $\langle\xi\rangle$ for the $U(1)_{B-L}^{}$ symmetry breaking, the heavy Higgs singlets $\sigma_a^{}$ or $\omega_b^{}$ can acquire a suppressed VEV, 
\begin{eqnarray}
\!\!\!\!\langle\sigma_a^{}\rangle &\simeq & -\frac{\kappa_{\sigma_a^{}}^{}\langle\xi\rangle^3_{}}{M_{\sigma_a^{}}^2}\ll \langle\xi\rangle~~\textrm{for}~~M_{\sigma_a^{}}^{}\gg \langle\xi\rangle\,,\\
\!\!\!\!\langle\omega_b^{}\rangle &\simeq & -\frac{\rho_{\omega_b^{}}^{}\langle\xi\rangle^2_{}}{M_{\omega_b^{}}^2}~~\textrm{for}~~M_{\omega_b^{}}^{}\gtrsim \rho_{\omega_b^{}}^{}\,,~M_{\omega_b^{}}^{}\gg \langle\xi\rangle\,.
\end{eqnarray}
In consequence, the second Higgs doublet $\eta$ can obtain a small mixing with the SM Higgs doublet $\phi$, i.e.
\begin{eqnarray}
\mathcal{L}&\supset& - \mu_{\eta\phi}^2 \eta^\dagger_{}\phi +\textrm{H.c.}~~\textrm{with}\nonumber\\
&&\mu_{\eta\phi}^2=\sum_{a}^{} \rho_{\sigma_a^{}}^{}\langle\sigma_a^{}\rangle  ~~\textrm{or}~~\mu_{\eta\phi}^2=\sum_{b}^{} \kappa_{\omega_b^{}}^{}\langle\omega_b^{}\rangle\langle\xi\rangle\,.\nonumber\\
&&
\end{eqnarray}
We can see the suppressed VEV $\langle\sigma_a^{}\rangle$ or $\langle\omega_b^{}\rangle$ and then small quadratic coupling $\mu_{\eta\phi}^{2}$ are induced by a seesaw mechanism like the traditional type-II seesaw \cite{mw1980}.

Subsequently, the SM Higgs doublet $\phi$ acquires a VEV $\langle\phi\rangle\simeq 174\,\textrm{GeV}$ for the electroweak symmetry breaking. The neutral and charged components of the second Higgs doublet $\eta=(\eta^0_{},~\eta^{-}_{})^T_{}$ have their masses as below,  
\begin{eqnarray}
m_{\eta^0_{}}^2&=&m_\eta^2 +\lambda_{\eta\xi}^{}\langle\xi\rangle^2_{}+ ( \lambda_{\eta\phi}^{}+ \lambda'^{}_{\eta\phi})\langle\phi\rangle^2_{}  \,,\nonumber\\
m_{\eta^{\pm}_{}}^2&=& m_\eta^2 +\lambda_{\eta\xi}^{}\langle\xi\rangle^2_{}+ \lambda_{\eta\phi}^{} \langle\phi\rangle^2_{} \,.
\end{eqnarray}
The neutral component $\eta^0_{}$ can pick up a small VEV, 
\begin{eqnarray}
\langle\eta\rangle\simeq -\frac{ \mu_{\eta\phi}^2\langle\phi\rangle}{m_{\eta^0_{}}^2} \ll \langle\phi\rangle~~\textrm{for}~~|\mu_{\eta\phi}^2|\ll m_{\eta^0_{}}^2\,.
\end{eqnarray}
The left-handed neutrinos $\nu_{L}^{}$ and the first and second right-handed neutrinos $\nu_{R1,2}^{}$ can naturally obtain a tiny Dirac mass term,
\begin{eqnarray}
\mathcal{L}&\supset& -\sum_{i=1,2 \atop \alpha=e,\mu,\tau}^{}\left(m_\nu^{}\right)_{\alpha i}^{} \bar{\nu}_{L\alpha}^{} \nu_{Ri}^{} +\textrm{H.c.}~~\textrm{with}~~m_\nu^{} = f \langle\eta\rangle\,.\nonumber\\
&&
\end{eqnarray} 
The above Dirac neutrino mass generation by the small VEV $\langle\eta\rangle$ is again a seesaw mechanism like the traditional type-II seesaw \cite{mw1980} for the Majorana neutrinos.

In conclusion, we have realized a two-step seesaw mechanism to generate the tiny Dirac neutrino masses. Such Dirac neutrino mass generation may be entitled as a double type-II Dirac seesaw \cite{gu2019-1}, in analogy to our double type-II seesaw \cite{ghsz2009} for the Majorana neutrinos. Note this Dirac neutrino mass matrix can only have two nonzero eigenvalues because it is a $3\times 2$ matrix with two right-handed neutrinos. In this sense, we may further name our neutrino mass generation as a minimal double type-II Dirac seesaw.

For a numerical estimation, we choose
\begin{eqnarray}
\langle\xi\rangle=\mathcal{O}(100\,\textrm{TeV})\,,
\end{eqnarray}
and
\begin{eqnarray}
&&M_{\sigma_a^{}/\omega_b^{}}^{}=\mathcal{O}(10^{16}_{}\,\textrm{GeV})\,,~~\rho_{\sigma_a^{}/\omega_b^{}}^{}=\mathcal{O}(10^{14}_{}\,\textrm{GeV})\,,\nonumber\\
&&\kappa_{\sigma_a^{}/\omega_b^{}}^{}=\mathcal{O}(0.01)\,.
\end{eqnarray}
We then obtain
\begin{eqnarray}
\!\!\!\!\!\!\!\!&&\mu_{\eta\phi}^2 =  \mathcal{O}(10^{-5}_{}\,\textrm{GeV}^2_{})~~\textrm{for}\nonumber\\
\!\!\!\!\!\!\!\!&&\langle\sigma_a^{}\rangle = \mathcal{O}(10^{-19}_{}\,\textrm{GeV})~~\textrm{or}~~\langle \omega_b^{} \rangle =\mathcal{O}(10^{-8}_{}\,\textrm{GeV})\,.
\end{eqnarray}
By further taking 
\begin{eqnarray}
m_{\eta^0_{}}^{}= \mathcal{O}(1-10\,\textrm{TeV})\,,
\end{eqnarray}
we realize
\begin{eqnarray}
\langle\eta\rangle = \mathcal{O}(0.01-1\,\textrm{eV})\,.
\end{eqnarray}
and hence 
\begin{eqnarray}
\!\!\!\!\!\!\!\!&&m_\nu^{}=\langle\eta\rangle = \mathcal{O}(0.01-0.1\,\textrm{eV})~~\textrm{for}~~f_{\alpha i}^{} =\mathcal{O}(0.1-1)\,.\nonumber\\
\!\!\!\!\!\!\!\!&&
\end{eqnarray}

\section{Baryon asymmetry}

\begin{figure*}
\vspace{7.5cm} \epsfig{file=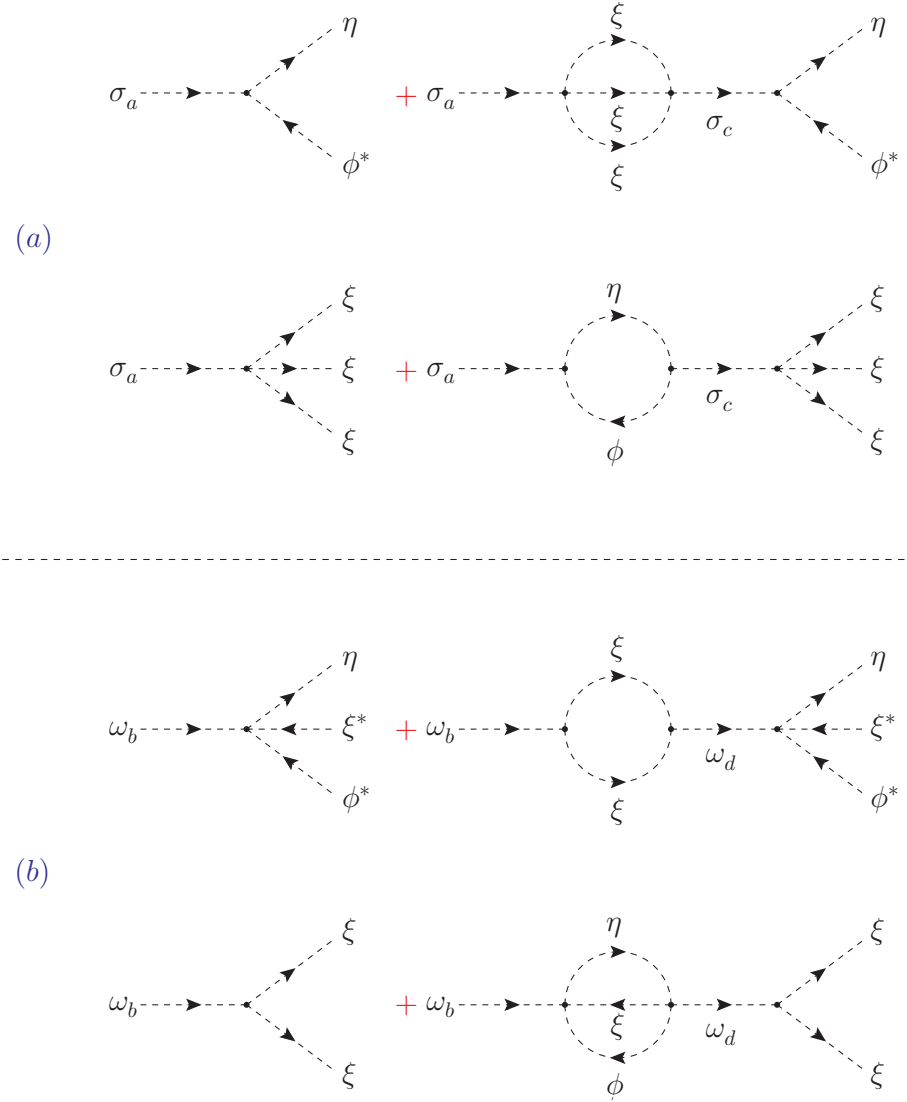, bbllx=5cm, bblly=6.0cm,
bburx=15cm, bbury=16cm, width=8cm, height=8cm, angle=0,
clip=0} \vspace{0cm} \caption{\label{sdecay} The heavy Higgs singlet decays.}
\end{figure*}

As shown in Fig. \ref{sdecay}, the heavy Higgs singlets $\sigma_a^{}$ have two decay modes,
\begin{eqnarray}
\sigma_a^{}\rightarrow \xi\xi\xi\,,~~ \sigma_a^{} \rightarrow \eta \phi^\ast_{}\,.
\end{eqnarray}
As long as the CP is not conserved, we can expect a CP asymmetry in the above decays, 
\begin{eqnarray}
\varepsilon_{\sigma_a}^{} &=& \frac{ \Gamma(\sigma_a^{} \rightarrow \eta \phi^\ast_{}) - \Gamma(\sigma_a^{\ast} \rightarrow \eta^\ast_{} \phi ) }{\Gamma_a^{}}\nonumber\\
&=&- \frac{\Gamma(\sigma_a^{}\rightarrow \xi\xi\xi) - \Gamma(\sigma_a^{\ast}\rightarrow \xi^\ast_{}\xi^\ast_{}\xi^\ast_{}) }{\Gamma_a^{}}\neq 0\,,\end{eqnarray}
where $\Gamma_a^{} $ is the total decay width,
\begin{eqnarray}
\Gamma_{\sigma_a}^{} &=& \Gamma(\sigma_a^{}\rightarrow \xi\xi\xi) + \Gamma(\sigma_a^{} \rightarrow \eta \phi^\ast_{}) \nonumber\\
&=&  \Gamma(\sigma_a^{\ast}\rightarrow \xi^\ast_{}\xi^\ast_{}\xi^\ast_{}) + \Gamma(\sigma_a^{\ast} \rightarrow \eta^\ast_{} \phi ) \,.
\end{eqnarray}
We can calculate the decay width at tree level and the CP asymmetry at one-loop level, 
\begin{eqnarray}
\Gamma_{\sigma_a}^{} &=&\frac{1}{8\pi}\left(\frac{|\rho_{\sigma_a}^{}|^2}{M_{\sigma_a}^2}+\frac{3|\kappa_{\sigma_a}^{}|^2_{}}{32\pi^2_{}}\right) M_{\sigma_a}^{}\,,
\end{eqnarray}
\begin{eqnarray}
\varepsilon_{\sigma_a}^{} &=&- \frac{3}{64\pi^3_{}}\sum_{c\neq a}^{}\frac{\textrm{Im}\left(\kappa^\ast_{\sigma_a}\kappa_{\sigma_c}^{}\rho_{\sigma_a}^{}\rho_{\sigma_c}^\ast\right)}{\frac{|\rho_{\sigma_a}^{}|^2}{M_{\sigma_a}^2}+\frac{3|\kappa_{\sigma_a}^{}|^2_{}}{32\pi^2_{}}}\frac{1}{M_{\sigma_c}^2-M_{\sigma_a}^2}\nonumber\\
&=&- \frac{3}{64\pi^3_{}}\sum_{c\neq a}^{}\frac{|\kappa^{}_{\sigma_a}\kappa_{\sigma_c}^{}\rho_{\sigma_a}^{}\rho_{\sigma_c}^{}|\sin\sigma_{ac}^{}}{\frac{|\rho_{\sigma_a}^{}|^2}{M_{\sigma_a}^2}+\frac{3|\kappa_{\sigma_a}^{}|^2_{}}{32\pi^2_{}}}\frac{1}{M_{\sigma_c}^2-M_{\sigma_a}^2}\,.\nonumber\\
&&
\end{eqnarray}
Here $\sigma_{ac}^{}$ is the relative phase among the parameters $\rho_{\sigma_a,\sigma_c}^{}$ and $\kappa_{\sigma_a,\sigma_c}^{}$.

Similarly, the heavy Higgs singlets $\omega_b^{}$ also have a two-body decay and a three-body decay,
\begin{eqnarray}
\omega_a^{}\rightarrow \xi\xi\,,~~ \omega_b^{} \rightarrow \eta \phi^\ast_{}\xi^\ast_{}\,.
\end{eqnarray}
We can calculate the decay width and the CP asymmetry,
\begin{eqnarray}
\Gamma_{\omega_b}^{} &=& \Gamma(\omega_b^{}\rightarrow \xi\xi) + \Gamma(\omega_b^{} \rightarrow \eta \phi^\ast_{}\xi^\ast_{}) \nonumber\\
&=&  \Gamma(\omega_b^{\ast}\rightarrow \xi^\ast_{}\xi^\ast_{}) + \Gamma(\omega_b^{\ast} \rightarrow \eta^\ast_{} \phi \xi ) \nonumber\\
&=&\frac{1}{8\pi}\left(\frac{|\rho_{\omega_b}^{}|^2}{M_{\omega_b}^2}+\frac{|\kappa_{\omega_b}^{}|^2_{}}{32\pi^2_{}}\right) M_{\omega_b}^{}\,,
\end{eqnarray}
\begin{eqnarray}
\varepsilon_{\omega_b}^{} &=& \frac{ \Gamma(\omega_b^{} \rightarrow \eta \phi^\ast_{}\xi^\ast_{}) - \Gamma(\omega_b^{\ast} \rightarrow \eta^\ast_{} \phi \xi) }{\Gamma_{\omega_b^{}}^{}}\nonumber\\
&=&- \frac{\Gamma(\omega_b^{}\rightarrow \xi\xi) - \Gamma(\omega_b^{\ast}\rightarrow \xi^\ast_{}\xi^\ast_{}) }{\Gamma_{\omega_b^{}}^{}}\nonumber\\
&=&- \frac{1}{64\pi^3_{}}\sum_{d\neq b}^{}\frac{\textrm{Im}\left(\kappa^\ast_{\omega_b}\kappa_{\omega_d}^{}\rho_{\omega_b}^{}\rho_{\omega_d}^\ast\right)}{\frac{|\rho_{\omega_b}^{}|^2}{M_{\omega_b}^2}+\frac{|\kappa_{\omega_b}^{}|^2_{}}{32\pi^2_{}}}\frac{1}{M_{\omega_d}^2-M_{\omega_b}^2}\nonumber\\
&=&- \frac{1}{64\pi^3_{}}\sum_{d\neq b}^{}\frac{|\kappa^{}_{\omega_b}\kappa_{\omega_c}^{}\rho_{\omega_b}^{}\rho_{\omega_d}^{}|\sin\omega_{ab}^{}}{\frac{|\rho_{\omega_b}^{}|^2}{M_{\omega_b}^2}+\frac{|\kappa_{\omega_b}^{}|^2_{}}{32\pi^2_{}}}\frac{1}{M_{\omega_d}^2-M_{\omega_b}^2}\,.\nonumber\\
&&
\end{eqnarray}
Here $\omega_{bd}^{}$ is the relative phase among the parameters $\rho_{\omega_b,\omega_d}^{}$ and $\kappa_{\omega_b,\omega_d}^{}$.

After the heavy Higgs singlets $\sigma_a^{}/\omega_b^{}$ go out of equilibrium, their decays can generate an asymmetry $A_\eta^{}$ stored in the second Higgs doublet $\eta$. For demonstration, we simply assume a hierarchical spectrum of the heavy Higgs singlets $\sigma_a^{}/\omega_b^{}$, i.e. $M_{\sigma_1^{}/\omega_1^{}}^{2}\ll M_{\sigma_{2,...}^{}/\omega_{2,...}^{}}^2$. In this case, the decays of the lightest $\sigma_1^{}$ should dominate the final $A_\eta^{}$ asymmetry, i.e.
\begin{eqnarray}
A_\eta^{}&=& \varepsilon_{\sigma_1^{}/\omega_1^{}}^{}\left(\frac{n^{eq}_{\sigma_1^{}/\omega_1^{}} }{s}\right)\left|_{T=T_D^{}}^{}\right.\,,
\end{eqnarray}
where the symbols $n^{eq}_{\sigma_1^{}/\omega_1^{}}$ and $T_D^{}$ respectively are the equilibrium number density and the decoupled temperature of the heavy Higgs singlets $\sigma_1^{}/\omega_1^{}$, while the character $s$ is the entropy density of the universe \cite{kt1990}. The sphaleron processes eventually will partially transfer this $A_\eta^{}$ asymmetry to a baryon asymmetry \cite{ht1990},
\begin{eqnarray}
\label{bauf}
B= -\frac{28}{79} \varepsilon_{\sigma_1^{}/\omega_1^{}}^{}\left(\frac{n^{eq}_{\sigma_1^{}/\omega_1^{}} }{s}\right)\left|_{T=T_D^{}}^{}\right.\,.
\end{eqnarray}

For a numerical example, we consider the case with two heavy Higgs singlets $\sigma_{1,2}^{}/\omega_{1,2}$\,. We then take 
\begin{eqnarray}
&&M_{\sigma_1/\omega_1}^{}=10^{16}_{}\,\textrm{GeV}\,,~|\rho_{\sigma_1/\omega_1}^{}|=10^{14}_{}\,\textrm{GeV}\,, \nonumber\\
&&~|\kappa_{\sigma_1/\omega_1}^{}|=0.02\,;~~M_{\sigma_2/\omega_2^{}}^{}=3\times 10^{16}_{}\,\textrm{GeV}\,,\nonumber\\
&&|\rho_{\sigma_2/\omega_2}^{}| =3\times 10^{14}_{}\,\textrm{GeV}\,,~|\kappa_{\sigma_2/\omega_2}^{}|=0.02\,,\nonumber\\
&&
\end{eqnarray}
and hence obtain
\begin{eqnarray}
\Gamma_{\sigma_1}^{}&=&4\times 10^{10}_{}\,,\textrm{GeV}\,,\nonumber\\
\varepsilon_{\sigma_1}^{}&=&-2.3\times 10^{-7} \sin\sigma_{12}^{}\,;~~\textrm{or}\nonumber\\
\Gamma_{\omega_1}^{}&= &4\times 10^{10}_{}\,\textrm{GeV}\,,\nonumber\\
\varepsilon_{\omega_1}^{}&=&-7.6\times 10^{-8} \sin\omega_{12}^{}\,.
\end{eqnarray}
We can see the decay width $\Gamma_{\sigma_1/\omega_1}^{}$ is smaller than the Hubble constant $H(T)$ at the temperature $T=M_{\sigma_1/\omega_1}^{}$, i.e.
\begin{eqnarray}
&&\left.\left[H(T)=\left(\frac{8\pi^{3}_{}g_{\ast}^{}}{90}\right)^{\frac{1}{2}}_{}\frac{T^2_{}}{M_{\textrm{Pl}}^{}}\right]\right|_{T=M_{\sigma_1^{}/\omega_1^{}}}^{} \nonumber\\
&&= 1.5\times 10^{14}_{}\,\textrm{GeV}\,,
\end{eqnarray}
with $M_{\textrm{Pl}}^{}\simeq 1.22\times 10^{19}_{}\,\textrm{GeV}$ being the Planck mass and $g_{\ast}^{}=121.75$ being the relativistic degrees of freedom (the SM fields plus the four right-handed neutrinos $\nu_{R}^{}$, the second Higgs doublet $\eta$, the Higgs singlet $\xi$ and the $U(1)_{B-L}^{}$ gauge boson.). Therefore, the weak washout condition \cite{kt1990},
\begin{eqnarray}
\frac{\Gamma_{\sigma_1/\omega_1}^{}}{H(T=M_{\sigma_1^{}/\omega_1^{}})}<1\,,
\end{eqnarray}
can be satisfied in the heavy Higgs singlet decays. The baryon asymmetry (\ref{bauf}) hence can be simply given by 
\begin{eqnarray}
\label{bauf}
B&\sim&  -\frac{28}{79} \frac{\varepsilon_{\sigma_1^{}}^{}}{g_\ast^{}}  =10^{-10}_{}\left(\frac{\sin\sigma_{12}^{}}{0.15}\right)~~\textrm{or}\nonumber\\
B&\sim&  -\frac{28}{79} \frac{\varepsilon_{\omega_1^{}}^{}}{g_\ast^{}}  = 10^{-10}_{}\left(\frac{\sin\sigma_{12}^{}}{0.45}\right)\,.
\end{eqnarray}

\section{Dark matter and right-handed neutrinos}

After the $U(1)_{B-L}^{}$ symmetry breaking, the gauge boson $Z_{B-L}^{}$ can obtain its mass 
\begin{eqnarray}
M_{Z_{B-L}^{}}^{}\simeq  \frac{\sqrt{2}}{5}g_{B-L}^{} \langle \xi\rangle\,.
\end{eqnarray}
Meanwhile, the third and forth right-handed neutrinos $\nu_{R3,4}^{}$ can form a Dirac particle \cite{pry2016,gu2019-2,gu2019-3}, i.e.
\begin{eqnarray}
\mathcal{L} &\supset&  i \bar{\chi}\gamma^\mu_{}\partial_\mu^{} \chi - m_\chi^{} \bar{\chi}\chi \nonumber\\
&& \textrm{with}~~ \chi = \nu_{R3}^{}+\nu_{R4}^{c}\,,~~ m_\chi^{}= y_{34}^{} \langle\xi\rangle\,.
 \end{eqnarray}
This Dirac fermion has no decay channels so that it can be a stable dark matter particle. The dark matter annihilation and scattering could be dominated by the gauge interactions,
\begin{eqnarray}
\mathcal{L} &\supset&  g_{B-L}^{} Z_{B-L}^\mu \left\{\sum_{i=1}^{3}\left(\frac{1}{3}\bar{d}_i^{}\gamma_\mu^{} d_i^{} + \frac{1}{3}\bar{u}_{i}^{}\gamma_\mu^{}u_{i}^{}-\bar{e}_{i}^{}\gamma_\mu^{} e_{i}^{} \right. \right.\nonumber\\
&&\left.\left.-\bar{\nu}_{Li}\gamma_\mu^{} \nu_{Li}^{}\right)-\frac{8}{5} \bar{\nu}_{R1}^{}\gamma_\mu^{}\nu_{R1}^{} - \frac{8}{5} \bar{\nu}_{R2}^{}\gamma_\mu^{}\nu_{R2}^{}\right.\nonumber\\
&&\left.-i\frac{3}{5}\left[\left(\partial_\mu^{}\eta\right)^\dagger_{}\eta - \textrm{H.c.}\right]  + \frac{1}{10}\bar{\chi}\gamma_\mu^{}\left(\sqrt{865}+\gamma_5^{}\right)\chi  \right\}\,.\nonumber\\
&&
\end{eqnarray}
The perturbation requirement then constrain the gauge coupling $g_{B-L}^{}$ by
\begin{eqnarray}
\label{gbl}
\frac{\sqrt{865}}{10}g_{B-L}^{} < \sqrt{4\pi} \Rightarrow g_{B-L}^{}<  \sqrt{\frac{80\pi}{173}}\,,
\end{eqnarray} 
while the experimental results constrain the $U(1)_{B-L}^{}$ symmetry breaking scale by \cite{afpr2017,klq2016}
\begin{eqnarray}
\label{low}
\frac{M_{Z_{B-L}^{}}^{}}{g_{B-L}^{} }  \gtrsim 7 \,\textrm{TeV} \Rightarrow \langle \xi \rangle \gtrsim 25\,\textrm{TeV}\,.
\end{eqnarray}

We can calculate the thermally averaging dark matter annihilating cross section \cite{bhkk2009},  
\begin{eqnarray}
\label{ann}
\langle\sigma_{\textrm{A}}^{} v_{\textrm{rel}}^{} \rangle&=& \sum_{f=d,u,e,\nu_{L}^{},\nu_{R1,2}^{}}^{}\langle\sigma(\chi+\chi^c_{}\rightarrow f+f^c_{}) v_{\textrm{rel}}^{}\rangle \nonumber\\
&&+ \langle\sigma(\chi+\chi^c_{}\rightarrow \eta+\eta^\ast_{}) v_{\textrm{rel}}^{}\rangle \nonumber\\
&\simeq &\frac{39963 g_{B-L}^4}{500\pi} \frac{m_\chi^2}{M_{Z_{B-L}^{}}^4} \nonumber\\
&=& \frac{199815}{16\pi} \frac{m_\chi^2}{\langle\xi\rangle^4_{}} =  \frac{199815}{16\pi} \frac{y_{\chi}^2}{\langle\xi\rangle^2_{}}  \,,
\end{eqnarray}
by simply assuming 
\begin{eqnarray}\label{ychi}
4m_\chi^2 \ll M_{Z_{B-L}^{}}^2& \Rightarrow& y_{\chi}^{2}\ll \frac{1}{50} g_{B-L}^{2} < \frac{8\pi}{865}\nonumber\\
&\Rightarrow & y_\chi^{}< \sqrt{\frac{8\pi}{865}}\,.
\end{eqnarray}
The dark matter relic density then can approximately given by \cite{pdg2018}
\begin{eqnarray}
\label{relic}
\Omega_\chi^{} h^2 \simeq \frac{0.1\,\textrm{pb}}{\langle \sigma_{\textrm{A}}^{} v_{\textrm{rel}}^{} \rangle} &=&0.1\,\textrm{pb}\times \frac{16\pi \langle\xi\rangle^4_{}}{199815m_{\chi}^2}\nonumber\\
&=&0.1\,\textrm{pb}\times \frac{16\pi \langle\xi\rangle^2_{}}{199815 y_{\chi}^2} \,.
\end{eqnarray}

From Eqs. (\ref{ychi}) and (\ref{relic}), we can put a constraint on the the VEV $\langle\xi\rangle$, i.e.
\begin{eqnarray}
\langle\xi\rangle &\simeq& \left(\frac{199815 y_{\chi}^2 \, \Omega_\chi^{} h^2}{16\pi \times  0.1\,\textrm{pb}} \right)^{\frac{1}{2}}_{}\nonumber\\
&=&  222\,\textrm{TeV}\left(\frac{y_{\chi}^{}}{\sqrt{8\pi/865}}\right) \left(\frac{\Omega_\chi^{} h^2}{0.11}\right)^{\frac{1}{2}}_{}\nonumber\\
&<&222\,\textrm{TeV}\left(\frac{\Omega_\chi^{} h^2}{0.11}\right)^{\frac{1}{2}}_{}\,.
\end{eqnarray}
besides the experimental limit (\ref{low}). The dark matter mass, 
\begin{eqnarray}
\label{dm1}
m_\chi^{} &\simeq &\left(0.1\,\textrm{pb}\times \frac{16\pi\langle\xi\rangle^4_{}}{199815 \Omega_\chi^{} h^2} \right)^{\frac{1}{2}}_{} \nonumber\\
&=& 5\,\textrm{TeV}\left(\frac{\langle \xi\rangle }{81\,\textrm{TeV}}\right)^2_{}\left(\frac{0.11}{\Omega_\chi^{} h^2}\right)^{\frac{1}{2}}_{} \,,
\end{eqnarray}
thus should be in the range, 
\begin{eqnarray}
\label{dm2}
&&480\,\textrm{GeV}\left(\frac{0.11}{\Omega_\chi^{} h^2}\right)^{\frac{1}{2}}_{}\lesssim
m_\chi^{} < 38\,\textrm{TeV}\left(\frac{0.11}{\Omega_\chi^{} h^2}\right)^{\frac{1}{2}}_{}\nonumber\\
&&\textrm{for}~~25\,\textrm{TeV} \lesssim \langle\xi\rangle < 222\,\textrm{TeV}\,.
\end{eqnarray}

The gauge interactions can also mediate the dark matter scattering off nucleons. The dominant scattering cross section is spin independent \cite{jkg1996}, 
\begin{eqnarray}
\sigma_{\chi N}^{}&=& \frac{173g_{B-L}^4}{20 \pi} \frac{\mu_r^2 }{M_{Z_{B-L}^{}}^4} \nonumber\\
&=& \frac{21625 }{16\pi} \frac{\mu_r^2 }{\langle\xi\rangle^4_{}} =   \frac{25}{231} \frac{\mu_r^2 }{m_\chi^2} \frac{0.1\,\textrm{pb}}{\Omega_\chi^{} h^2_{} }\,.
\end{eqnarray}
Here $\mu_r^{}=m_N^{}m_\chi^{}/(m_N^{}+m_\chi^{})$ is a reduced mass with $m_N^{}$ being the nucleon mass. As the dark matter is much heavier than the nucleon, the above dark matter scattering cross section indeed should be inversely proportional to the squared dark matter mass, 
\begin{eqnarray}
\sigma_{\chi N}^{}&=& 3.5\times 10^{-45}_{}\,\textrm{cm}^2_{}\left(\frac{\mu_r^{}}{940\,\textrm{MeV}}\right)^2_{} \left(\frac{0.11}{\Omega_\chi^{} h^2_{}}\right)\nonumber\\
&&\times \left( \frac{5\,\textrm{TeV}}{m_\chi^{}}\right)^2_{} \,.
\end{eqnarray}
The dark matter direct detection results \cite{cui2017,aprile2018} then put a low limit on the dark matter mass, i.e. 
\begin{eqnarray}
\label{dm3}
m_\chi^{} \gtrsim 5\,\textrm{TeV}\,.
\end{eqnarray}
So, the dark matter masses (\ref{dm2}) should be modified by
\begin{eqnarray}
\label{dm3}
&&5\,\textrm{TeV}\left(\frac{0.11}{\Omega_\chi^{} h^2}\right)^{\frac{1}{2}}_{}\lesssim
m_\chi^{} < 38\,\textrm{TeV}\left(\frac{0.11}{\Omega_\chi^{} h^2}\right)^{\frac{1}{2}}_{}\nonumber\\
&&\textrm{for}~~81\,\textrm{TeV} \lesssim \langle\xi\rangle < 222\,\textrm{TeV}\,.
\end{eqnarray}

The two right-handed neutrinos $\nu_{R1,2}^{}$ would significantly contribute to the effective neutrino number \cite{pdg2018} if they did not decouple above the QCD scale. This is not allowed by the BBN. So, we check the annihilations of the right-handed neutrinos into the relativistic species at the QCD scale,   
\begin{eqnarray}
\sigma_{\nu_R^{}}^{} &=&\sum_{f=d,u,s,e,\mu,\nu_L^{}}^{}\sigma(\nu_R^{}+\nu_R^c\rightarrow f+f^c) \nonumber\\
&=& \frac{24 g_{B-L}^4}{25\pi}\frac{s}{M_{Z_{B-L}}^4} =  \frac{150}{\pi}\frac{s}{\langle\xi\rangle^4_{}}\,,
\end{eqnarray}
with $s$ being the Mandelstam variable. The interaction rate then should be \cite{gnrrs2003}
\begin{eqnarray}
\Gamma_{\nu_R^{}}^{} =\frac{\frac{T}{32\pi^4_{}}\int^{\infty}_{0} s^{3/2}_{} K_1^{}\left(\frac{\sqrt{s}}{T}\right) \sigma_{\zeta}^{}ds }{\frac{2}{\pi^2_{}}T^3_{}}= \frac{1800}{\pi^3_{}} \frac{T^5_{}}{\langle\xi\rangle^4_{}}\,,
\end{eqnarray}
with $K_1^{}$ being a Bessel function. We take $g_\ast^{}(300\,\textrm{MeV})\simeq 61.75$ and then find 
\begin{eqnarray}
\left[\Gamma_{\nu_R^{}}^{} < H(T)\right]_{T\gtrsim 300\,\textrm{MeV}}^{}~~\textrm{for}~~\langle\xi\rangle \gtrsim 51\,\textrm{TeV}\,,
\end{eqnarray}
to fulfil the parameter space (\ref{dm3}).

\section{Conclusion}

In this paper we have shown the $U(1)_{B-L}^{}$ gauge symmetry with four neutral fermions can predict two right-handed neutrinos for the Dirac neutrino mass generation and one stable Dirac fermion for the dark matter relic density. Specifically, two neutral fermions can form a Dirac fermion for a stable dark matter particle through their Yukawa couplings to the Higgs singlet for spontaneously breaking the $U(1)_{B-L}^{}$ symmetry. The other two neutral fermions as the right-handed neutrinos can couple to the SM lepton doublets with a second Higgs doublet. After the $U(1)_{B-L}^{}$ symmetry breaking, some heavy Higgs singlets can acquire their suppressed VEVs to highly suppress the mixing between the second Higgs doublet and the SM Higgs doublet. Consequently, the second Higgs doublet can naturally pick up a tiny VEV even if it is at the TeV scale. This means a testable Dirac neutrino mass generation. Remarkably, the neutrino mass matrix can only have two nonzero eigenvalues since it just involoves two right-handed neutrinos. The interactions for generating the neutrino masses can also explain the observed baryon asymmetry in association with the sphaleron processes.

\textbf{Acknowledgement}: This work was supported by the National Natural Science Foundation of China under Grant No. 11675100 and the Recruitment Program for Young Professionals under Grant No. 15Z127060004.

\appendix

\section{The $U(1)_{B-L}^{}$ gauge anomalies }

The $SU(3)_c^{}-SU(3)_c^{}-U(1)_{B-L}^{}$ anomaly is
\begin{eqnarray}
\!\!\!\!&&3\times 3\times \left[2\times \left(+\frac{1}{3}\right) -\left(+ \frac{1}{3}\right) - \left(+\frac{1}{3}\right) \right]=0\,.
\end{eqnarray}

The $SU(2)_L^{}-SU(2)_L^{}-U(1)_{B-L}^{}$ anomaly is 
\begin{eqnarray}
3\times 2 \times  \left[3\times\left(+ \frac{1}{3}\right) +\left(-1\right)\right]=0\,.
\end{eqnarray}

The $U(1)_Y^{}-U(1)_Y^{}-U(1)_{B-L}^{}$ anomaly is 
\begin{eqnarray}
\!\!\!\!&&3\times \left\{3\times \left[2\times \left(+\frac{1}{6}\right)^{\!2}_{} - \left(-\frac{1}{3}\right)^{\!2}_{} - \left(+\frac{2}{3}\right)^{\!2}_{}\right] \times \left(+ \frac{1}{3}\right) \right.\nonumber\\
[2mm]
\!\!\!\!&&\left.+\left[2\times \left(-\frac{1}{2}\right)^2_{} - \left(-1\right)^2_{} \right]\times \left(-1\right) \right\}  =0\,.
\end{eqnarray}

The $U(1)_Y^{}-U(1)_{B-L}^{}-U(1)_{B-L}^{}$ anomaly is 
\begin{eqnarray}
&&3\times \left\{3\times \left[2\times \left(+\frac{1}{6}\right) -\left(-\frac{1}{3}\right)-\left(+\frac{2}{3}\right)\right]\times \left(+\frac{1}{3}\right)^{\!2}_{} \right.\nonumber\\
[2mm]
&&\left.+\left[2\times \left(-\frac{1}{2}\right) -\left(-1\right)\right]\times \left(-1\right) \right\}=0\,.
\end{eqnarray}

The $U(1)_{B-L}^{}-U(1)_{B-L}^{}-U(1)_{B-L}^{}$ anomaly is 
\begin{eqnarray}
&&3\times \left\{3\times \left[2\times \left(+\frac{1}{3}\right)^{\!3}_{}-\left(+\frac{1}{3}\right)^{\!3}_{}-\left(+\frac{1}{3}\right)^{\!3}_{}\right] \right.\nonumber\\
[2mm]
&&\left.+\left[2\times \left(-1\right)^{3}_{} -\left(-1\right)^{3}_{}\right] \right\} -2\times \left(-\frac{8}{5}\right)^{3}_{}  \nonumber\\
[2mm]
&&-\left(\frac{1-\sqrt{865}}{10}\right)^{\!3}_{} -\left(\frac{1+\sqrt{865}}{10}\right)^{\!3}_{}=0\,.
\end{eqnarray}

The graviton-graviton-$U(1)_{B-L}^{}$ anomaly is
\begin{eqnarray}
&&3\times \left\{3\times \left[2\times \left(+\frac{1}{3}\right)-\left(+\frac{1}{3}\right)-\left(+\frac{1}{3}\right)\right] \right.\nonumber\\
[2mm]
&&\left.+\left[2\times \left(-1\right) -\left(-1\right)\right] \right\} -2\times \left(-\frac{8}{5}\right) \nonumber\\
[2mm]
&&-\left(\frac{1-\sqrt{865}}{10}\right) -\left(\frac{1+\sqrt{865}}{10}\right)=0\,.
\end{eqnarray}

\end{document}